\documentclass[11pt]{article}\def\today{March 13, 2006}
\usepackage{amssymb}
\usepackage{amsmath}
\usepackage{theorem}
\usepackage{latexsym}
\usepackage{graphicx}
\reversemarginpar
\theoremheaderfont{\bf} 
\theorembodyfont{\sl}
\newtheorem{theo}{Theorem}[section]
{\theorembodyfont{\rm} \newtheorem{defi}[theo]{Definition}}
{\theorembodyfont{\rm} \newtheorem{exa}[theo]{Example}}
{\theorembodyfont{\rm} \newtheorem{rem}[theo]{Remark}}
\newtheorem{prop}[theo]{Proposition}

{\theorembodyfont{\rm} }
\newtheorem{lemma}[theo]{Lemma}
{\theorembodyfont{\rm}}
{\theorembodyfont{\rm}}
\newenvironment{proof}{{\sc Proof:}}{\mbox{}\hfill$\Box$\par}
\newcommand{\eqnref}[1]{~\mbox{$(${\rm \ref{#1}}$)$}}

\newcommand{\junk}[1]{}

\newcommand{\N}{{\mathbb N}}
\newcommand{\F}{{\mathbb F}}
\newcommand{\Z}{{\mathbb Z}}
\newcommand{\cC}{{\mathcal C}}

\newcommand{\rank}{\mbox{\rm rk}\,}

\newcommand{\McM}{\mbox{$\delta_{\text{M}}$}}
\newcommand{\AutF}{\mbox{$\text{Aut}(\F)$}}

\newcommand{\im}{\mbox{\rm im}\,}

\newcommand{\wt}{\mbox{\rm wt}}
\newcommand{\we}{\mbox{\rm we}}

\newcommand{\T}{\mbox{$\!^{\sf T}$}}

\newcommand{\Smallfourmat}[4]{\mbox{\tiny{$\begin{pmatrix}{#1}&{\!\!\!#2}\\{#3}&{\!\!\!#4}\end{pmatrix}$}}}
\newenvironment{liste}{\begin{list}{--\hfill}{\topsep0ex \labelwidth.4cm
   \leftmargin.5cm \labelsep.1cm \rightmargin0cm \parsep0ex \itemsep.6ex
   \partopsep1.4ex}}{\end{list}}
\newcounter{abc}
\newcounter{def}
\newenvironment{romanlist}{\begin{list}{(\roman{abc})\hfill}{\usecounter{abc}
     \topsep-1.4ex \labelwidth.7cm \leftmargin.7cm \labelsep0cm
     \rightmargin0cm \parsep0ex \itemsep.6ex
     \partopsep1.6ex}}{\end{list}}
\newenvironment{alphalist}{\begin{list}{(\alph{abc})\hfill}{\usecounter{abc}
     \topsep0ex \labelwidth.7cm \leftmargin.7cm \labelsep0cm
     \rightmargin0cm \parsep0ex \itemsep.6ex
     \partopsep1.6ex}}{\end{list}}
\newenvironment{arabiclist}{\begin{list}{(\arabic{abc})\hfill}{\usecounter{abc}
     \topsep-1.4ex \labelwidth.7cm \leftmargin.7cm \labelsep0cm
     \rightmargin0cm \parsep0ex \itemsep.6ex
     \partopsep1.6ex}}{\end{list}}

\title{State Space Realizations and Monomial Equivalence for Convolutional Codes}
\date\today
\author{Heide Gluesing-Luerssen$^*$, Gert Schneider\footnote{
       University of Groningen, Department of Mathematics, P.~O.~Box 800,
       9700 AV Groningen, The Netherlands; gluesing@math.rug.nl, schneider@math.rug.nl}
       }

\begin{document}
\maketitle

{\bf Abstract:} We will study convolutional codes with the help of state space realizations.
It will be shown that two such minimal realizations belong to the same code if and only if they are equivalent under the full state feedback group.
This result will be used in order to prove that two codes with positive Forney indices are monomially equivalent if and only if they share the same adjacency matrix.
The adjacency matrix counts in a detailed way the weights of all possible outputs and thus contains full information about the weights of the codewords in the given code.

{\bf Keywords:} Convolutional codes, minimal realizations, weight adjacency matrix, monomial equivalence

{\bf MSC (2000):} 94B10, 94B05, 93B15, 93B20

\section{Introduction}
\setcounter{equation}{0}

In the theory of linear block codes the Equivalence Theorem of MacWilliams~\cite{MacW62,MacW63} tells us that two block codes are isometric if and only if they are monomially equivalent, that is, if they differ only by permutation and rescaling of the coordinates.
In other words, the intrinsic notion of isometry coincides with the extrinsically defined concept of  monomial equivalence.
This theorem became the cornerstone of the notion of equivalence for block codes and allows us to classify these codes.
Since the discovery of the importance of linear block codes over~$\Z_4$ for nonlinear codes, the Equivalence Theorem has enjoyed various generalizations to block codes over certain finite rings,
see for instance the articles \cite{WaWo96,Wo99a,GrSch00,DiLP04a}.

For convolutional codes a classification, taking all relevant parameters of the code into account,
has not yet been established.
In other words, it is not yet clear as to when two such codes should be identified.
In this paper we want to contribute to this issue by showing that the adjacency matrix forms a complete invariant under monomial equivalence for convolutional codes with positive Forney indices.

The adjacency matrix of a code counts in a very detailed and systematic way the weights of codeword coefficients.
It will be introduced in Section~3, and its properties, as found in~\cite{GL05p,GS06}, will be briefly summarized.
All that will indicate that it forms an adequate generalization of the classical weight enumerator for block codes.
The adjacency matrix is defined via suitable state space realizations of reduced encoders.
In this sense, our approach follows a series of papers where convolutional codes have been investigated successfully by system-theoretic methods, see, e.~g., \cite{RoYo99,Ro01,HRS05}.
Since for a given code neither the reduced encoders nor the associated realizations are
unique, we will first discuss in detail the relationship between any two minimal realizations
for a given code.
This is accomplished in Section~2 by making use of classical realization theory.
It turns out that, in essence, two minimal realizations belong to the same code if and only if
they are equivalent under the full state feedback group.
In Section~3 the weight adjacency matrix associated with a minimal realization will be introduced.
The only-if part of the theorem just mentioned will provide us with an easy way to turn this matrix into an invariant of the code.
Finally, the if-part together with MacWilliams' Equivalence Theorem for block codes will lead to our main result stating that two convolutional codes with positive Forney indices are  monomially equivalent if and only if they share the same adjacency matrix.
This result is not true for codes where at least one Forney index is zero; in particular it is not true for block codes (of dimension bigger than one), which, of course, is a well known fact.

We strongly believe that our main theorem will be helpful in order to establish an appropriate notion of equivalence, and thus a useful classification, for convolutional codes.
It should be clear that a reasonable notion of equivalence should involve  those (vector space) isometries that leave all error-correcting properties of the code invariant.
Since the adjacency matrix comprises many of the parameters characterizing these properties~\cite[Sec.~3]{GL05p} we believe that it is reasonable to require this matrix to be invariant under code equivalence.
In this sense our main theorem can be regarded as a generalization of MacWilliams' Equivalence Theorem to convolutional codes with positive Forney indices.
However, the result does not tell us how in general an adequate notion of code equivalence should look like, and we have to leave this open for future research.

We end the introduction with recalling the basic notions for convolutional codes as needed in this paper.
Let~$\F$ be a finite field.
A {\sl $k$-dimensional convolutional code of length\/} $n$ is a submodule~$\cC$ of $\F[z]^n$ of the form
\[
    \cC=\im G:=\{uG\,\big|\, u\in\F[z]^k\}
\]
where~$G$ is a {\sl basic\/} matrix in $\F[z]^{k\times n}$, i.~e.
\begin{equation}\label{e-Gbasic}
  \rank G(\lambda)=k \text{ for all }\lambda\in\overline{\F},
\end{equation}
with $\overline{\F}$ being an algebraic closure of~$\F$.
In other words,~$G$ is noncatastrophic and delay-free.
We call such a matrix $G$ an {\sl encoder}, and the number
\begin{equation}\label{e-degG}
   \deg(\cC):=\deg(G):=\max\{\deg(M)\mid M\text{ is a $k$-minor of }G\}
\end{equation}
is said to be the {\sl degree\/} of the encoder~$G$ or of the code~$\cC$.
A matrix $G\in\F[z]^{k\times n}$ with rows $g_1,\ldots, g_k\in\F[z]^n$ is said to be
{\sl reduced\/} if $\sum_{i=1}^k\deg (g_i)=\deg(G)$.
Here $\deg(g_i)$ denotes the $i$th row degree, i.~e., the maximal degree of the entries in the
$i$th row of~$G$.
For characterizations of reducedness see, e.~g., \cite[Main~Thm.]{Fo75} or \cite[Thm.~A.2]{McE98}.
It is well known~\cite[p.~495]{Fo75} that each convolutional code~$\cC$ admits a reduced encoder~$G$.
The row degrees $\deg(g_i)$ of a reduced encoder~$G$ are, up to ordering, uniquely determined by the code and are called the {\sl Forney indices\/} of the code or of the encoder.
It follows that a convolutional code has a constant encoder matrix if and only if the degree is zero.
In that case the code is, in a natural way, a block code.

\section{State Space Descriptions of Reduced Encoders}\label{S-ABCD}
\setcounter{equation}{0}
In this section we will study state space descriptions of convolutional codes and discuss their non-uniqueness.
More precisely, we will fix a code $\cC=\im G$ and concentrate on the encoding process
\begin{equation}\label{e-uG}
   G:\,\F[z]^k\longrightarrow\cC,\quad u\longmapsto v:=uG
\end{equation}
for various (reduced) encoders $G\in\F[z]^{k\times n}$.
Obviously, the encoding\eqnref{e-uG} can be interpreted as a dynamical input-output system and
thus can be described as a state space system in the system theoretic sense.
In this section we will describe all possible state space descriptions of a given code~$\cC$ with
minimal state space dimension and investigate their relation to
each other.
The main difficulty will be the non-uniqueness of the encoder matrix~$G$.
The considerations of this section are, of course, closely related to classical realization theory, and can be deduced straightforwardly. 
However, the polynomial rather than proper rational setting and the fact that not the encoder but rather the code is the object under consideration lead to certain differences, and we consider it worth
deriving the results in detail. 
Of course, we will make use of classical realization theory.
It is worth mentioning that the results of this section are true for arbitrary fields~$\F$ and do not require the finiteness of~$\F$.

In order to use standard notation of systems theory it will be most convenient to associate
the proper rational transfer matrix
\begin{equation}\label{e-TG}
  T_G(z):=G(z^{-1})\in\F(z)^{k\times n}
\end{equation}
to a polynomial matrix $G\in\F[z]^{k\times n}$.
Notice that the transfer function~$T_G$ is polynomial in~$z^{-1}$, or, in other words, $T_G$ does not have any poles in $\overline{\F}\backslash\{0\}$.
Recall that the {\sl McMillan degree\/} $\McM(T)$ of a proper rational matrix $T\in\F(z)^{k\times n}$ can be defined as $\McM(T):=\deg(\det Q)$ where
\begin{equation}\label{e-Tcoprime}
   T=Q^{-1}P\,\text{ is a coprime factorization with matrices }
   Q\in\F[z]^{k\times k},\; P\in\F[z]^{k\times n}.
\end{equation}
Coprimeness of the factorization $Q^{-1}P$ simply means that the matrix $[Q,P]$ is basic.
It is well known that such a factorization always exists (e.~g., the Smith-McMillan form), and the McMillan degree does not depend on the choice of the coprime factorization.

\begin{prop}\label{P-GT}
Let $G\in\F[z]^{k\times n}$ be a polynomial matrix and let $T_G$ be as in\eqnref{e-TG}.
Then $\McM(T_G)\geq\deg(G)$.
Moreover, if~$G$ is reduced then $\McM(T_G)=\deg(G)$.
\end{prop}
\begin{proof}
Let $\McM(T_G)=\gamma$ and $\deg(G)=\delta$.
Since all possible poles of~$T_G$ are at zero we know that~$T_G$ has a factorization as in\eqnref{e-Tcoprime}
such that $\det Q=z^{\gamma}$.
Let $m:=\sum_{i=0}^\delta m_iz^i$ be a $k$-minor of~$G$ of degree~$\delta$.
Then $P(z)=Q(z)G(z^{-1})$ being polynomial implies that
$z^{\gamma}m(z^{-1})=\sum_{i=0}^\delta m_iz^{\gamma-i}$ is polynomial. Since $m_{\delta}\not=0$ this yields $\gamma\geq\delta$.
\\
Let now $G$ be reduced and $\nu_1,\ldots,\nu_k$ be the row degrees of~$G$.
Moreover, for $i=1,\ldots,k$ let the $i$th row of~$G$ be $g_i=\sum_{l=0}^{\nu_i}g_{i,l}z^l$ where $g_{i,l}\in\F^n$.
Then
\begin{equation}\label{e-copfac}
  T_G(z)=\text{diag}(z^{\nu_1},\ldots,z^{\nu_k})^{-1}
       \Big(\sum_{l=0}^{\nu_i}g_{i,\nu_i-l}z^l\Big)_{i=1,\ldots,k}.
\end{equation}
Now reducedness of~$G$ implies that the rightmost matrix has full row rank for $z=0$, see \cite[Main~Thm.]{Fo75}.
This in turn yields that\eqnref{e-copfac} is a coprime polynomial factorization of~$T_G$, and we
obtain the desired result.
\end{proof}

The last statement of Proposition~\ref{P-GT} is not an if-and-only-if statement.
This can easily be verified using the matrix in\eqnref{e-Gexa} below.

Let us now turn to state space realizations of encoders.
Before we briefly recall some well known results from realization theory as to be found, e.~g., in \cite[Ch.~6]{Kail80}, let us remind that transfer matrices act on input vectors from the right, see\eqnref{e-uG}.
Taking the according notational changes into account we obtain that each proper rational matrix $T\in\F(z)^{k\times n}$ has a {\sl realization\/} $(A,B,C,D)\in\F^{\delta\times\delta+k\times\delta+\delta\times n+k\times n}$, meaning that $T(z)=B(zI-A)^{-1}C+D$.
Furthermore, $\delta\geq\McM(T)$, and $\delta=\McM(T)$ if and only if $(A,B,C,D)$ is {\sl controllable\/} and {\sl observable}, that is,
\[
   \rank\begin{pmatrix}B\\BA\\ BA^2\\ \vdots\end{pmatrix}=\delta \ \text{ and }\
   \rank\begin{pmatrix}C&AC&A^2C&\ldots\end{pmatrix}=\delta.
\]
Controllable and observable realizations for a given $T\in\F(z)^{k\times n}$ do always exist.
They are unique up to {\sl similarity}; precisely, given two such realizations
$(A,B,C,D)$ and $(\bar{A},\bar{B},\bar{C},\bar{D})$ of~$T$ then there exists a matrix
$S\in GL_{\delta}(\F)$ such that
\begin{equation}\label{e-sim}
   (\bar{A},\bar{B},\bar{C},\bar{D})=(SAS^{-1},BS^{-1},SC,D).
\end{equation}

Assume now that $T=T_G$ for some matrix $G\in\F[z]^{k\times n}$ with full row rank.
Thus the only poles (if any) of the rational matrix~$T$ are at zero.
One can show straightforwardly that any realization~$(A,B,C,D)$ of~$T$ leads to the equivalence
\begin{equation}\label{e-iso}
  v=uG
        \Longleftrightarrow
        \left\{\begin{array}{rcl} x_{t+1}&=&x_tA+u_tB\\v_t&=&x_tC+u_tD\end{array}
          \;\text{ for all }t\geq0\right\} \text{ where }x_0=0
\end{equation}
for any $u=\sum_{t\geq0}u_tz^t\in\F[z]^k$ and $v=\sum_{t\geq0}v_tz^t\in\F[z]^n$,
see also \cite[Thm.~2.3]{GL05p}.
Indeed, the state vectors~$x_t$ are given as the coefficients of the polynomial $x=uB(z^{-1}I-A)^{-1}$.
Due to this interpretation we simply call the quadruple $(A,B,C,D)$ a {\sl (state space) system\/} over~$\F$.
The discussion gives rise to the following definition.
\begin{defi}\label{D-real}
Let $(A,B,C,D)\in\F^{\delta\times\delta+k\times\delta+\delta\times n+k\times n}$ be a system over~$\F$.
\begin{arabiclist}
\item Let $G\in\F[z]^{k\times n}$ be a polynomial matrix with full row rank.
      Then $(A,B,C,D)$ is said to be a {\sl realization of order\/} $\delta$ of~$G$ if
      \[
          G(z)=B(z^{-1}I-A)^{-1}C+D.
      \]
      As usual, the system is called {\sl canonical\/} if it is controllable and observable.
\item We call $(A,B,C,D)$ a {\sl realization of the code\/} $\cC\subseteq\F[z]^n$
      if there exists an encoder $G\in\F[z]^{k\times n}$ of~$\cC$ such that $(A,B,C,D)$ is a realization of~$G$.
      If~$G$ is reduced and $(A,B,C,D)$ is a canonical realization of~$G$, then it is said to be a {\sl canonical minimal realization\/} of~$\cC$.
\end{arabiclist}
\end{defi}

Since a realization of~$G$ is, by definition, a realization of the proper matrix $T_G$ in the system theoretic sense, it follows from the discussion above that each polynomial matrix~$G$ has a realization, and the order of any realization is at least $\McM\big(T_G\big)$.
Each such~$G$ also has a canonical realization, and a given realization is canonical if and only if its
order equals $\McM\big(T_G\big)$.
Any two canonical realizations $(A,B,C,D)$ and $(\bar{A},\bar{B},\bar{C},\bar{D})$ of~$G$ are {\sl similar\/} in the sense of\eqnref{e-sim}.
Moreover, due to Proposition~\ref{P-GT} a realization of a reduced matrix~$G$ is canonical if and
only if it has order $\deg(G)$.
Finally, each code has a canonical minimal realization; it has order $\deg(\cC)$.
Let us also note that in the special case where $\deg(\cC)=0$, i.~e.,~$\cC$ is a block code, the
matrices $A,B,C$ of a canonical minimal realization do not exist and $D=G$, where~$G$ is a constant encoder of~$\cC$.

We will single out a particularly simple realization of a given encoder.
\begin{prop}\label{P-CCF}
Let $G\in\F[z]^{k\times n}$ be a polynomial matrix with rank~$k$ and row degrees
$\nu_1,\,\ldots,\nu_k$.
Put $\delta:=\sum_{i=1}^k\nu_i$.
Let~$G$ have rows $g_i=\sum_{\ell=0}^{\nu_i}g_{i,\ell}z^{\ell},\,i=1,\ldots,k,$ where
$g_{i,\ell}\in\F^n$.
For $i=1,\ldots,k$ define the matrices
\[
 A_i=\left(\begin{smallmatrix} 0&1& & \\ & &\ddots& \\& & &1\\ & & &0\end{smallmatrix}\right)
      \in\F^{\nu_i\times\nu_i},\
 B_i=\begin{pmatrix}1&0&\cdots&0\end{pmatrix}\in\F^{\nu_i},\
 C_i=\begin{pmatrix}g_{i,1}\\ \vdots\\ g_{i,\nu_i}\end{pmatrix}\in\F^{\nu_i\times n}.
\]
Then the {\sl controller form\/} of~$G$ is defined as
the matrix quadruple
$(A,B,C,D)\in\F^{\delta\times\delta}\times\F^{k\times\delta}\times
             \F^{\delta\times n}\times\F^{k\times n}$
where
\[
   A=\left(\begin{smallmatrix} A_1&  & \\ &\ddots &\\ & &A_k\end{smallmatrix}\right),\:
   B=\left(\begin{smallmatrix}
            B_1& &\\ &\ddots & \\ & &B_k\end{smallmatrix}\right),\:
   C=\left(\begin{smallmatrix}C_1\\ \vdots\\[.5ex]C_k\end{smallmatrix}\right),\:
   D=\left(\begin{smallmatrix}g_{1,0}\\[.4ex] \vdots\\[.6ex]g_{k,0}\end{smallmatrix}\right)=G(0).
\]
In the case where $\nu_i=0$ the $i$th block is missing and in~$B$ a zero row occurs.
The following is true.
\begin{romanlist}
\item The controller form $(A,B,C,D)$ forms a controllable realization of the matrix~$G$.
\item $G$ is reduced if and only if $\rank\Smallfourmat{-A}{C}{-B}{D}=\delta+k$.
\item If~$G$ is reduced, then the controller form is a canonical realization.
\end{romanlist}
\end{prop}
\begin{proof}
Part~(i) is proved in \cite[Prop.~2.1]{GL05p} and part~(ii) can be checked
directly\footnote{The equivalence given in \cite[(2.2)]{GL05p} is false in general.
It is true, however, if all row degrees of~$G$ are positive.}.
Part~(iii) is a consequence of~(i) and~(ii) since observability is equivalent to $\rank(\lambda I-A,C)=\delta$ for all $\lambda\in\overline{\F}$ which, due to nilpotency of~$A$, is equivalent to
$\rank(-A,C)=\delta$.
But the latter follows from~(ii).
\end{proof}

It is well known, and can also straightforwardly be shown, that if~$G$ is reduced the controller form is the shift realization of the
coprime factorization\eqnref{e-copfac} of~$T_G$ as introduced and discussed in detail by Fuhrmann~\cite[Thm.~10-1]{Fu81}.

We will now investigate those canonical systems $(A,B,C,D)$ that give rise to a
polynomial basic and reduced encoder matrix $G(z)=B(z^{-1}I-A)^{-1}C+D$.

\begin{theo}\label{T-ABCDmain}
Let $(A,B,C,D)\in\F^{\delta\times\delta+k\times\delta+\delta\times n+k\times n}$ be a canonical system
and put $G:=B(z^{-1}I-A)^{-1}C+D\in\F(z)^{k\times n}$.
Then $G$ is a polynomial matrix if and only if~$A$ is nilpotent.
If $A$ is nilpotent one also has the following.
\begin{alphalist}
\item $G$ is a basic polynomial matrix if and only if $\rank D=k$ and
      $\rank\Smallfourmat{\lambda I-A}{C}{-B}{D}=\delta+k$ for all
      $\lambda\in\overline{\F}\backslash\{0\}$.
\item If $G$ is a reduced polynomial matrix then
      $\rank\Smallfourmat{-A}{C}{-B}{D}=\delta+k$.
\end{alphalist}
Summarizing, if $G$ is a basic and reduced polynomial matrix then
\begin{equation}\label{e-cond}
   A \text{ nilpotent,}\quad
   \rank D=k, \quad
   \rank\begin{pmatrix}\lambda I-A&C\\ -B&D\end{pmatrix}=\delta+k
   \text{ for all }\lambda\in\overline{\F}.
\end{equation}
\end{theo}
\begin{proof}
For the first statement notice that $G=D+\sum_{i=1}^{\infty}BA^{i-1}Cz^i$.
This shows immediately the if-part.
On the other hand, if~$G$ is polynomial then there exists an index $N\in\N$ such that $BA^iC=0$ for $i\geq N$.
Thus
\[
  0=\begin{pmatrix}B\\BA\\BA^2\\ \vdots\end{pmatrix}
    \begin{pmatrix}A^NC&A^{N+1}C&A^{N+2}C&\cdots\end{pmatrix}
  =\begin{pmatrix}B\\BA\\BA^2\\ \vdots\end{pmatrix}A^N
  \begin{pmatrix}C&AC&A^{2}C&\cdots\end{pmatrix}.
\]
Now controllability and observability yield $A^N=0$ as desired.
\\[.6ex]
(a)
By nilpotency of~$A$ the matrix $\lambda I-A$ is regular for $\lambda\not=0$.
Thus
\[
  \rank\begin{pmatrix}\lambda I-A&C\\ -B&D\end{pmatrix}
  =\rank\begin{pmatrix}\lambda I-A&C\\ 0&D+B(\lambda I-A)^{-1}C\end{pmatrix}
  =\rank\begin{pmatrix}\lambda I-A&C\\ 0& G(\lambda^{-1})\end{pmatrix}.
\]
Since $G(0)=D$ this yields that~$G$ is basic iff
$\rank\Smallfourmat{\lambda I-A}{C}{-B}{D}=\delta+k$ for all $\lambda\in\overline{\F}\backslash\{0\}$ and $\rank D=k$.
\\[.6ex]
(b) Let $G$ be reduced and consider the controller form $(A,B,C,D)$ of~$G$.
Then the required rank condition is satisfied by Proposition~\ref{P-CCF}(ii).
By part~(iii) of that proposition the controller form is canonical.
Now~(b) follows for arbitrary canonical realizations by using the facts that each such realization is similar to the controller form and that the rank
of $\Smallfourmat{\lambda I-A}{C}{-B}{D}$ is invariant under similarity.
\end{proof}

\begin{rem}\label{R-zeros}
Using the transformation $T=T_G$ part~(a) of the last theorem is a particular instance of the well-known system theoretic fact that the transmission zeros (i.~e., the zeros of the transfer matrix) coincide with the invariant zeros of a canonical realization (i.~e., the zeros of the rightmost matrix in\eqnref{e-cond}), see for instance \cite[p.~578]{Kail80}.
Part~(b) reflects the fact that row reduced matrices have no zeros at infinity, see \cite[6.5.-19, p.~468]{Kail80}.
Indeed, by definition~$G$ has a zero at infinity if~$T_G$ has a zero at zero, meaning that $\rank P(0)<k$ for a coprime factorization $Q^{-1}P=T_G$.
But if~$G$ is reduced then the factorization in\eqnref{e-copfac} shows that~$T_G$ has no zeros at zero.
The converse of Theorem~\ref{T-ABCDmain}(b), and thus the converse of this last statement, is not true.
This can be seen from the system
\[
  \begin{pmatrix}zI-A&C\\-B&D\end{pmatrix}
  =\left(\!
   \begin{array}{c|ccc}z&0&0&1\\\hline 1&0&1&1\\2&1&0&0\end{array}\!\right)
   \in\F_3[z]^{3\times 4}.
\]
The system is controllable and observable and satisfies\eqnref{e-cond}.
But the corresponding polynomial matrix
\begin{equation}\label{e-Gexa}
   G=B(z^{-1}I-A)^{-1}C+D=\begin{pmatrix}0&1&1+2z\\1&0&z\end{pmatrix}
\end{equation}
is not reduced.
As a consequence, $(A,B,C,D)$ is not a canonical minimal realization of the code $\cC=\im G$ in the
sense of Definition~\ref{D-real}.
It can also be checked straightforwardly that the matrix~$T_G$ has no zeros at zero.
\end{rem}

Notice that any system satisfying\eqnref{e-cond} is observable (but not necessarily controllable).
It should also be observed that conversely, even if the first two conditions of\eqnref{e-cond} are satisfied and the system is controllable,
observability does not imply the last condition of\eqnref{e-cond}.
For instance, the system over~$\F_2$ given by
\[
  A=\begin{pmatrix}0&1\\0&0\end{pmatrix},\
  B=\begin{pmatrix}1&0\end{pmatrix},\ C=\begin{pmatrix}0&1\\1&0\end{pmatrix},\
  D=\begin{pmatrix}1&1\end{pmatrix}
\]
is controllable and observable and satisfies the first two conditions of\eqnref{e-cond}, but not
the last one.
Its encoder matrix is given by $G=\begin{pmatrix}1+z&1+z^2\end{pmatrix}$ which is not basic.

Let us now turn to canonical realizations of different reduced encoders for a given code.
The following lemma will be crucial.
\begin{lemma}\label{L-feedback}
Let $(A,B,C,D)\in\F^{\delta\times\delta+k\times\delta+\delta\times n+k\times n}$ satisfy\eqnref{e-cond}.
Furthermore, let $M\in\F^{\delta\times k}$ and put
\begin{equation}\label{e-feedback}
  \bar{A}=A-MB,\ \bar{B}=B,\ \bar{C}=C-MD,\ \bar{D}=D.
\end{equation}
Suppose that~$\bar{A}$ is nilpotent. Then
\begin{alphalist}
\item $(\bar{A},\bar{B},\bar{C},\bar{D})$ satisfies\eqnref{e-cond},
\item $B(z^{-1}I-A)^{-1}C+D=V(z)\big(\bar{B}(z^{-1}I-\bar{A})^{-1}\bar{C}+\bar{D}\big)$ where
      $V(z):=I+B(z^{-1}I-A)^{-1}M$,
\item $V\in GL_k(\F[z])$.
\end{alphalist}
\end{lemma}
\begin{proof}
(a) It suffices to observe that
\[
  \begin{pmatrix}\lambda I-\bar{A}&\bar{C}\\ -\bar{B}&\bar{D}\end{pmatrix}
  =\begin{pmatrix}I&-M\\ 0&I\end{pmatrix}
  \begin{pmatrix}\lambda I-A&C\\ -B&D\end{pmatrix}.
\]
(b) First notice that
$B(z^{-1}I-A+MB)^{-1}=\big(I+B(z^{-1}I-A)^{-1}M\big)^{-1}B(z^{-1}I-A)^{-1}$.
Putting $G(z):=B(z^{-1}I-A)^{-1}C+D$ and
$\bar{G}(z):=\bar{B}(z^{-1}I-\bar{A})^{-1}\bar{C}+\bar{D}$ we compute
\begin{align*}
  &V(z)^{-1}G(z)=\big(I+B(z^{-1}I-A)^{-1}M\big)^{-1}\big(B(z^{-1}I-A)^{-1}C+D\big)\\
  &\ =\big(I+B(z^{-1}I-A)^{-1}M\big)^{-1}\big(B(z^{-1}I-A)^{-1}(C-MD)+B(z^{-1}I-A)^{-1}MD+D\big)\\
  &\ =B(z^{-1}I\!-\!A\!+\!MB)^{-1}(C\!-\!MD)+\big(I\!+\!B(z^{-1}I\!-\!A)^{-1}M\big)^{-1}
      \big(B(z^{-1}I\!-\!A)^{-1}MD\!+\!D\big)\\
  &\ =B(z^{-1}I-A+MB)^{-1}(C-MD)+D\\
  &\ =\bar{G}(z).
\end{align*}
(c) Observe that $V\in\F[z]^{k\times k}$ by nilpotency of~$A$.
Moreover, $\det V\not\equiv 0$.
From Theorem~\ref{T-ABCDmain} we know that~$G$ and~$\bar{G}$ are both polynomial and basic.
Hence the unimodularity of~$V$ follows from~(b).
\end{proof}

One should observe that Identity~(b) above does not require nilpotency of~$A$.
After transformation $z\mapsto z^{-1}$, this identity for the transfer functions of state feedback equivalent systems is also known from systems theory and can in particular cases be deduced from,
for instance, \cite[Sec.~7.2.1]{Kail80}.
Our very particular situation makes~$V$ even a unimodular polynomial matrix.

Now we can prove the main result of this section.
\begin{theo}\label{T-feedbackequiv}
Let $G,\,\bar{G}\in\F[z]^{k\times n}$ be basic and reduced and let
$\deg(G)=\deg(\bar{G})=\delta$.
Let $(A,B,C,D)$ and $(\bar{A},\bar{B},\bar{C},\bar{D})$ be associated canonical realizations, respectively.
Then the following are equivalent.
\begin{romanlist}
\item $G=W\bar{G}$ for some $W\in GL_k(\F[z])$.
\item The systems $(A,B,C,D)$ and $(\bar{A},\bar{B},\bar{C},\bar{D})$ are equivalent under the full state feedback group, that is,
      there exist matrices $T\in GL_{\delta}(\F),\, U\in GL_k(\F),\, M\in\F^{\delta\times k}$
      such that
      \begin{equation}\label{e-ABCDtransform}
           \bar{A}=T^{-1}(A-MB)T,\ \bar{B}=UBT,\ \bar{C}=T^{-1}(C-MD),\ \bar{D}=UD.
      \end{equation}
\end{romanlist}
\end{theo}
\begin{proof}
(ii)~$\Rightarrow$~(i):
By Theorem~\ref{T-ABCDmain} both realizations satisfy\eqnref{e-cond}.
Hence we may apply Lemma~\ref{L-feedback}(b) and compute
\begin{align*}
  \bar{G}&=U\big(BT(z^{-1}I-T^{-1}(A-MB)T)^{-1}T^{-1}(C-MD)+D\big)\\
         &=U\big(B(z^{-1}I-A+MB)^{-1}(C-MD)+D\big)
         =UV^{-1}(z)\big(B(z^{-1}I-A)^{-1}C+D\big)
\end{align*}
where $V(z):=I+B(z^{-1}I-A)^{-1}M$. Now Lemma~\ref{L-feedback}(c) yields~(i).
\\
(i)~$\Rightarrow$~(ii):
First notice that equivalence under the full state feedback group is indeed an equivalence relation.
Since the controller form of a reduced matrix is canonical and all canonical realizations are similar, we may assume without
loss of generality that both $(A,B,C,D)$ and $(\bar{A},\bar{B},\bar{C},\bar{D})$
are in controller form.
Assumption~(i) implies that~$G$ and~$\bar{G}$ have the same row degrees.
Since reordering of the rows of~$G$ corresponds to a transformation $(T^{-1}AT,UBT,T^{-1}C,D)$ that retains the specific form of
the controller form we may further assume that $G$ and~$\bar{G}$ both have row degrees $\nu_1\geq\ldots\geq\nu_k$.
Then $A=\bar{A}$ and $B=\bar{B}$ since they are both fully determined by the row degrees.
Due to reducedness of~$G$ and~$\bar{G}$ the $i$th row of~$W$ has degree at most~$\nu_i$ for $i=1,\ldots,k$, see \cite[Main Thm.~(4)]{Fo75}.
We will show now that
\begin{equation}\label{e-Wreal}
  W=\big(I+B(z^{-1}I-A)^{-1}M\big)U^{-1}\text{ for some }
  M\in\F^{\delta\times k},\ U\in GL_{k}(\F).
\end{equation}
Then $U:=W(0)^{-1}$.
Hence we need to find~$M$ such that $B(z^{-1}I-A)^{-1}M=WU-I$.
The latter matrix is of the form $WU-I=\Big(\sum_{j=1}^{\nu_i}a_{ij}z^j\Big)_{i=1,\ldots,k}$ for suitable
$a_{ij}\in\F^k$.
Using that
\[
   B(z^{-1}I-A)^{-1}=
   \left(\!\!\begin{array}{ccccccccccccc}
      z&z^2&\cdots&z^{\nu_1}& &   &      &            &      &  &  &       &    \\
       &   &      &            &z&z^2&\cdots&z^{\nu_2}&      &  &  &       &    \\
       &   &      &            & &   &      &            &\ddots&  &  &       & \\
       &   &      &            & &   &      &            &\qquad&z &z^2&\cdots&z^{\nu_k}
   \end{array}\!\!\right),
\]
one sees that the matrix $M=(M_1,\ldots, M_k)\T$ where $M_i=(a_{i1}\T,\ldots,a_{i\nu_i}\T)$,
satisfies\eqnref{e-Wreal}.
Notice that if $\nu_i=0$ the result is true as well since in that case the $i$th block of~$M$ is missing
and a zero row appears in $WU-I$ and $B(z^{-1}I-A)^{-1}$.
Now we have the identity $G=\big(I+B(z^{-1}I-A)^{-1}M\big)U^{-1}\bar{G}$ which in turn implies
\[
  U\big(I+B(z^{-1}I-A)^{-1}M\big)^{-1}\big(B(z^{-1}I-A)^{-1}C+D\big)=B(z^{-1}I-A)^{-1}\bar{C}+\bar{D}.
\]
Using Lemma~\ref{L-feedback}(b) this yields
\begin{equation}\label{e-transfer}
  UB(z^{-1}I-A+MB)^{-1}(C-MD)+UD=B(z^{-1}I-A)^{-1}\bar{C}+\bar{D}=\bar{G}(z).
\end{equation}
Hence $(A-MB,UB,C-MD,UD)$ is a realization of $\bar{G}$ of order $\deg(\bar{G})$ and therefore
canonical.
As a consequence,\eqnref{e-transfer} implies that the realizations $(A-MB,UB,C-MD,UD)$ and $(A,B,\bar{C},\bar{D})$ are
similar, and this yields~(ii).
\end{proof}

Recall the notion of a canonical minimal realization of a code as given in Definition~\ref{D-real}(2).
The result just proven tells us that two canonical minimal realizations of a given code are equivalent under the full state feedback group.
The next example, however, shows that the action of the full state feedback group does in general not preserve the
property of being canonical minimal.

\begin{exa}\label{E-canmin}
Let $\F=\F_2$ and
\[
   A=\begin{pmatrix}0&0\\0&0\end{pmatrix},\   B=\begin{pmatrix}1&0\\0&1\end{pmatrix},\
   C=\begin{pmatrix}0&1&1\\0&0&1\end{pmatrix},\  D=\begin{pmatrix}1&0&1\\0&1&0\end{pmatrix}.
\]
Then $(A,B,C,D)$ is canonical and satisfies\eqnref{e-cond}.
Moreover,
\[
  G=B(z^{-1}I-A)^{-1}C+D=\begin{pmatrix}1&z&1+z\\0&1&z\end{pmatrix}\text{ is basic and reduced.}
\]
Thus, $(A,B,C,D)$ is a canonical minimal realization of the code $\cC=\im G$.
Using the feedback $M=\Smallfourmat{0}{0}{1}{0}$ and $T=U=I_2$ the system $(\bar{A},\bar{B},\bar{C},\bar{D})$
in\eqnref{e-ABCDtransform} leads to a nilpotent matrix~$\bar{A}$ and a non-reduced encoder matrix
\[
   \bar{G}=\bar{B}(z^{-1}I-\bar{A})^{-1}\bar{C}+\bar{D}
   =\begin{pmatrix}1&z&1+z\\z&1+z^2&z^2\end{pmatrix}.
\]
Hence the realization $(\bar{A},\bar{B},\bar{C},\bar{D})$ of the code~$\cC$ is not canonical minimal
in the sense of Definition~\ref{D-real}(2).
\end{exa}

The last example and Proposition~\ref{P-GT} suggest that the requirement of
reducedness for encoders seems too strong for our considerations.
Indeed, the results of this section become somewhat smoother if we replace reducedness by semi-reducedness where we call a matrix $G\in\F[z]^{k\times n}$ {\sl semi-reduced\/} if
$\McM(T_G)=\deg(G)$.
Let us briefly sketch the situation based on this notion.
\begin{liste}
\item By Proposition~\ref{P-GT} reducedness implies semi-reducedness,
      and obviously a matrix~$G$ is semi-reduced if and only if each canonical realization
      has order $\deg(G)$.
\item One can show that~$G$ is semi-reduced if and only if~$T_G$ has no zeros at zero.
      Recall that, by definition, the latter means $\rank P(0)=k$ for any coprime factorization as in\eqnref{e-Tcoprime}.
\item Using that the zeros of~$T_G$ coincide with the invariant zeros of a canonical
      realization \cite[p.~578]{Kail80} one obtains that Theorem~\ref{T-ABCDmain}(b) becomes an if-and-only if statement if we replace reducedness by semi-reducedness.
      As a consequence, a canonical realization $(A,B,C,D)$ satisfies\eqnref{e-cond} if and only if
      $G=B(z^{-1}I-A)^{-1}C+D$ is a basic and semi-reduced polynomial matrix.
\item Even Theorem~\ref{T-feedbackequiv} remains valid if we replace reducedness by semi-reducedness.
      Let us briefly outline the idea of the proof.
      The proof of the direction (ii)~$\Rightarrow$~(i) does not change at all.
      For the converse direction one first brings~$G$ into reduced form $\hat{W}G=\hat{G}$ where
      $\hat{W}\in GL_k(\F[z])$.
      Then it is not hard to see that, due to semi-reducedness of~$G$, the $i$th row degree of~$\hat{W}$ is less or equal the $i$th row degree of~$\hat{G}$.
      This allows us to proceed as in the proof of Theorem~\ref{T-feedbackequiv} and to show that $(A,B,C,D)$ is equivalent under the full feedback group to the controller form of~$\hat{G}$.
\item Finally, one can easily see that each state feedback action as in\eqnref{e-ABCDtransform},
      where~$M$ is such that $A-MB$ is nilpotent, transforms a canonical realization $(A,B,C,D)$
      satisfying\eqnref{e-cond} into a canonical system $(\bar{A},\bar{B},\bar{C},\bar{D})$ that also satisfies\eqnref{e-cond}.
      As a consequence, semi-reducedness of the corresponding encoder is preserved under each such transformation.
\end{liste}
Despite these more elegant results we decided to base this section on reducedness rather than semi-reducedness since firstly,
reducedness is a well-established concept and suffices for the results
needed here, and secondly, the proof of Theorem~\ref{T-feedbackequiv} makes explicit use of reducedness in either case.
Furthermore, in the next section we will need again the controller form of a reduced encoder.

\section{The Weight Adjacency Matrix and Monomial Equivalence}\label{S-MonEqu}
\setcounter{equation}{0}
In this section we will return to the particular situation of convolutional codes as
dynamical systems over finite fields.
Thus from now on let
\begin{equation}\label{e-Fq}
   \F=\F_q\text{ be a finite field with~$q=p^s$ elements where $p$ is prime and $s\in\N$}.
\end{equation}
Before introducing weight counting invariants for convolutional codes let us recall
the {\sl Hamming weight\/} on~$\F^n$ defined as
$\wt(w_1,\ldots,w_n):=\#\{i\mid w_i\not=0\}$.
We will also need the {\sl weight enumerator\/} of sets $S\subseteq \F^n$ given as
\begin{equation}\label{e-we}
  \we(S):=\sum_{i=0}^n \lambda_iW^i\in\Z[W], \text{ where }
  \lambda_i:=\#\{v\in S\mid \wt(v)=i\}.
\end{equation}
The weight enumerator $\we(\cC)$ of a block code $\cC\subseteq\F^n$ has been investigated intensively in the block coding literature.
For instance, the famous MacWilliams Identity Theorem~\cite{MacW63} tells us how to completely derive
$\we(\cC^{\perp})$ from $\we(\cC)$, where $\cC^{\perp}$ is the dual of~$\cC$ with
respect to the standard inner product on~$\F^n$.

The weight of a polynomial vector is defined straightforwardly by
simply extending the Hamming weight via
$\wt\big(\sum_{j=0}^Nv^{(j)}z^j\big):=\sum_{j=0}^N\wt(v^{(j)})$ for any
$v^{(j)}\in\F^n$.

In order to introduce an appropriate generalization of the weight enumerator for convolutional codes we need
the weight adjacency matrix of a realization.
Recall the state space system\eqnref{e-iso}.
It has $q^{\delta}$ different state vectors $x_t$ where~$\delta$ is the order of the realization $(A,B,C,D)$.
We consider now for each pair of states $(X,Y)\in\F^{2\delta}$ all (finitely many) state transitions
from $x_t=X$ to~$x_{t+1}=Y$ via suitable input $u_t=u$ and count
the weights of all corresponding outputs $v=XC+uD$.
This leads to the following definition, see also \cite[Sec.~2]{McE98a} and \cite[Def.~3.4]{GL05p}.
\begin{defi}\label{D-Lambda}
Let $G\in\F[z]^{k\times n}$ be a basic and reduced matrix such that $\deg(G)=\delta$ and let
$(A,B,C,D)$ be a canonical realization of~$G$.
We call $\F^{\delta}$ the {\sl state space\/} of the realization.
The {\sl weight adjacency matrix associated with\/} $(A,B,C,D)$ is defined to be the matrix
$\Lambda\in\Z[W]^{q^{\delta}\times q^{\delta}}$ that is
indexed by $(X,Y)\in\F^{2\delta}$ and has the entries
\begin{equation}\label{e-LambdaXY}
     \Lambda_{X,Y}:=\we\{XC+uD\mid u\in\F^k: Y=XA+uB\}\in\Z[W]\text{ for }(X,Y)\in\F^{2\delta}.
\end{equation}
\end{defi}

Recall that in the case $\delta=0$ the matrices $A,\,B,\,C$ do not exist while $D=G$.
As a consequence, $\Lambda=\Lambda_{0,0}=\we(\cC)$ is the ordinary weight enumerator $\we(\cC)$
of the block code $\cC=\{uG\mid u\in\F^k\}\subseteq\F^n$.

The properties of the weight adjacency matrix have been studied in detail in the papers~\cite{GL05p}
and~\cite{GS06}.
Among other things it has been discussed in detail in \cite[Sec.~3]{GL05p} that the weight adjacency matrix contains full
information about the extended row distances and the active burst distances of the convolutional code $\cC=\im G$.
These parameters are closely related to the error-correcting performance of~$\cC$ and are studied intensively in the more
engineering-oriented literature, see, e.~g.,~\cite{JPB90,HJZ02}.
In the paper~\cite{GS06}, alternative formulas for the entries of the weight adjacency matrix are given.
They are used in order to formulate a conjecture for a MacWilliams Identity for convolutional codes and their duals which
then is proven in special cases.
All this makes sense only because the weight adjacency matrix can indeed nicely be turned into an
invariant of the code. This will be shown below.
The discussion at this point should have made clear that this invariant will form an appropriate
generalization of the weight enumerator of block codes.

\begin{exa}\label{E-ExaLambda}
Let
\[
  G=\begin{pmatrix}z&1+z^2&1+z&z+z^2\\ 1&0&1&1\end{pmatrix}\in\F_2[z]^{2\times4}.
\]
Then $G$ is basic and reduced and the controller form is given by
\[
  A=\begin{pmatrix}0&1\\0&0\end{pmatrix},\
  B=\begin{pmatrix}1&0\\0&0\end{pmatrix},\
  C=\begin{pmatrix}1&0&1&1\\0&1&0&1\end{pmatrix},\
  D=\begin{pmatrix}0&1&1&0\\1&0&1&1\end{pmatrix}.
\]
In order to explicitly display the weight adjacency matrix we need to fix an ordering on the state space.
Let us choose the lexicographic ordering, hence $X_1=(0,0),\,X_2=(0,1),\,X_3=(1,0),\,X_4=(1,1)$.
Going through all possible combinations of states~$X$ and inputs~$u$ one obtains the weight
adjacency matrix
\[
   \Lambda=\begin{pmatrix}1+W^3&0&W^2+W^3&0\\W^2+W^3&0&W+W^2&0\\
                          0&1+W^3&0&W^2+W^3\\0&W^2+W^3&0&W+W^2\end{pmatrix},
\]
where the entry at position $(i,j)$ is $\Lambda_{X_i,X_j}$ as defined in\eqnref{e-LambdaXY}.
\end{exa}

The weight adjacency matrix does not form an invariant of a code but rather depends on the choice of both the reduced encoder and the canonical realization.
This dependence, however, can nicely be described.
\begin{theo}\label{T-Lambdaunique}
Let $\cC\subseteq\F[z]^n$ be a code of degree~$\delta$, and let
$(A,B,C,D)$ and $(\bar{A},\bar{B},\bar{C},\bar{D})$ both be canonical minimal realizations of~$\cC$.
Furthermore, let~$\Lambda$ and~$\bar{\Lambda}$ be the associated weight adjacency matrices, respectively.
Then there exists a state space isomorphism $T\in GL_{\delta}(\F)$ such that
\begin{equation}\label{e-Lambdatrafo}
  \bar{\Lambda}_{X,Y}=\Lambda_{XT,YT}\text{ for all } (X,Y)\in\F^{2\delta}.
\end{equation}
In particular, $\bar{\Lambda}=P\Lambda P^{-1}$ for some permutation matrix $P\in GL_{q^{\delta}}(\Z)$.
\end{theo}

The result appeared first in \cite[Remark~3.6, Theorem~4.1]{GL05p}.
Using Theorem~\ref{T-feedbackequiv} we can give an alternative, very short proof for this theorem.
Indeed, by Theorem~\ref{T-feedbackequiv} the two realizations are equivalent
under the full feedback group, thus we may assume\eqnref{e-ABCDtransform}.
But then one can straightforwardly check that for any $(X,Y,u,v)\in\F^{2\delta+k+n}$
\[
  Y=XA+uB,\ v=XC+uD
\]
is equivalent to
\[
  YT=XT\bar{A}+(uU^{-1}+XMU^{-1})\bar{B},\ v=XT\bar{C}+(uU^{-1}+XMU^{-1})\bar{D}.
\]
Since for any given~$X$ the mapping $u\longmapsto uU^{-1}+XMU^{-1}$ is bijective on~$\F^k$,  Equation\eqnref{e-Lambdatrafo} is immediate from the definition of the weight adjacency matrix.

The result above shows that we obtain an invariant of the code after factoring out the effect
of the state space isomorphism~$T$.
We will introduce this invariant in Definition~\ref{D-Lambdaequiv}(b) below in a slightly more general setting by also considering field automorphisms.
Thereafter we will investigate the relation between two codes having the same invariant.

Let us now turn to transformations on codes that obviously leave all relevant properties invariant.
It should be clear that field automorphisms are of this kind.
Likewise codes that differ only by a permutation and rescaling of the coordinates have the same characteristics.
In the sequel we will make these notions precise and show the effect of such transformations on the realizations and weight adjacency matrices.

Let $\AutF$ be the group of field automorphisms of~$\F=\F_{p^s}$.
Recall that each such automorphism leaves the prime field~$\F_p$ of~$\F$ invariant and thus is $\F_p$-linear.
For any $\phi\in\AutF$ we define its extension to polynomial matrices coefficientwise, that is,
\[
  \phi:\F[z]^{a\times b}\longrightarrow\F[z]^{a\times b},\quad
  M:=\Big(\sum_{\nu\geq0}m_{ij}^{(\nu)}z^{\nu}\Big)_{i,j}\longmapsto
  \phi(M):=\Big(\sum_{\nu\geq0}\phi(m_{ij}^{(\nu)})z^{\nu}\Big)_{i,j}.
\]
\begin{rem}\label{R-phi}
The mapping $\phi$ is multiplicative and $\F_p$-linear for matrices of fitting sizes
and satisfies $\rank\big(\phi(M)\big)=\rank(M)$.
Moreover, $\wt(v)=\wt\big(\phi(v)\big)$ for all $v\in\F[z]^n$.
Thus,~$\phi$ induces a weight-preserving $\F_p$-iso\-morphism on~$\F[z]^n$.
\end{rem}
Now we are ready to introduce monomial equivalence of convolutional codes.

\begin{defi}\label{D-me}
We define two matrices $G,\,G'\in\F[z]^{k\times n}$ with rank~$k$ to be {\sl monomially equivalent\/}
if there exists an automorphism $\phi\in\AutF$, a permutation matrix
$P\in GL_n(\F)$, and a nonsingular diagonal matrix $R\in GL_n(\F)$ such that $G'=\phi(G)PR$.
We call two codes {\sl monomially equivalent\/} if they have monomially equivalent encoder matrices.
\end{defi}
Obviously, monomial equivalence is an equivalence relation.
Moreover, monomially equivalent codes have the same dimension, Forney indices, and degree.
Furthermore, according to Remark~\ref{R-phi} the mapping $uG\longmapsto \phi(u)\phi(G)$ is weight-preserving and $\F_p$-linear.
Hence the same is true for $uG\longmapsto \phi(u)\phi(G)PR$, and thus
monomially equivalent codes are $\F_p$-isometric.
The isometry is even degree-preserving.
It should also be observed that, in general, testing whether two codes $\cC=\im G$ and $\cC'=\im G'$
of the same size are monomially equivalent can be quite a formidable task. 
Indeed, one has to check whether there exists a unimodular matrix~$U$, a permutation~$P$, and a diagonal matrix~$R$ such that $G'=UGPR$.

As we will see next, the effect of monomial equivalence on canonical realizations is
easily described.

\begin{prop}\label{P-meLambda}
Let the data be as in Definition~\ref{D-me} and assume $G'=\phi(G)PR$.
Let $\Sigma=(A,B,C,D)$ be any canonical realization of~$G$.
Then $\Sigma'=(\phi(A),\phi(B),\phi(C)PR,\phi(D)PR)$ is a canonical realization of~$G'$.
If~$G$, and thus~$G'$, is basic and reduced and $\Lambda,\,\Lambda'\in\Z[W]^{q^{\delta}\times q^{\delta}}$ are the weight adjacency matrices
associated with~$\Sigma$ and~$\Sigma'$, respectively, then
\begin{equation}\label{e-lambdatrafo}
      \Lambda_{X,Y}=\Lambda'_{\phi(X),\phi(Y)}\text{ for all }(X,Y)\in\F^{2\delta}.
\end{equation}
\end{prop}
\begin{proof}
Using multiplicativity and additivity of~$\phi$ one easily checks that~$\Sigma'$ is a canonical realization of~$G'$.
For the second part of the statement notice that for any $(X,Y)\in\F^{2\delta}$ and any $(u,v)\in\F^k\times\F^n$
one has
\begin{align*}
  Y=XA+uB&\Longleftrightarrow \phi(Y)=\phi(X)\phi(A)+\phi(u)\phi(B),\\
  v=XC+uD&\Longleftrightarrow \phi(v)PR=\phi(X)\phi(C)PR+\phi(u)\phi(D)PR.
\end{align*}
Since $\wt\big(\phi(v)PR\big)=\wt(v)$ for all $v\in\F^n$ the result follows directly from
Definition~\ref{D-Lambda}.
\end{proof}

The identities\eqnref{e-Lambdatrafo} and\eqnref{e-lambdatrafo} suggest the following equivalence
relation for weight adjacency matrices.
The corresponding equivalence classes result in the desired weight counting invariants for
convolutional codes.

\begin{defi}\label{D-Lambdaequiv}
\begin{alphalist}
\item For matrices $M,\,M'\in\Z[W]^{q^{\delta}\times q^{\delta}}$ we define the equivalence relation
    \[
    M\simeq M'\Longleftrightarrow \exists\,\phi\in\AutF,\,T\in GL_{\delta}(\F):\,
    M'_{X,Y}=M_{\phi(X)T,\phi(Y)T}
    \;\text{for all}\;(X,Y)\in\F^{2\delta}.
    \]
    We denote the equivalence class of~$M$ by
    $\overline{M}:=\{M'\in\Z[W]^{q^{\delta}\times q^{\delta}}\mid M'\simeq M\}$.
\item For a code~$\cC$ of degree~$\delta$ let $\bar{\Lambda}(\cC):=\overline{\Lambda}$ be the equivalence
      class of the adjacency matrix~$\Lambda$ associated with any canonical minimal realization
      of~$\cC$.
      We call $\bar{\Lambda}(\cC)$ the {\sl adjacency matrix\/} of the code~$\cC$.
\end{alphalist}
\end{defi}

It is worth mentioning that~$\simeq$ is not identical to the equivalence relation
in~\cite[(4.1)]{GL05p} since the latter does not cover $\F$-automorphisms.

From Theorem~\ref{T-Lambdaunique} we know that $\bar{\Lambda}(\cC)$ is indeed an invariant of the code.
Furthermore, Proposition~\ref{P-meLambda} implies that~$\bar{\Lambda}(\cC)$ is even invariant under monomial equivalence.
The main result of this section states that under a certain condition on the Forney indices the adjacency matrix even forms a {\sl complete\/} invariant for monomial equivalence.
Indeed, we have the following.
\begin{theo}\label{T-me}
Let $\cC,\,\cC'\subseteq\F[z]^n$ be two codes and assume that all Forney indices of~$\cC$ are positive.
Then $\cC$ and~$\cC'$ are monomially equivalent if and only if $\bar{\Lambda}(\cC)=\bar{\Lambda}(\cC')$.
\end{theo}

Notice that we require that~$\cC$ and~$\cC'$ are defined over the same field~$\F$ and have the same length~$n$.
Just like in block coding theory we consider this a reasonable assumption for this kind of considerations.
In the proof we will see that if $\bar{\Lambda}(\cC)=\bar{\Lambda}(\cC')$ the codes~$\cC$ and~$\cC'$ have the same Forney indices.
Thus the assumption above on the Forney indices is true for~$\cC'$ as well.

Remembering that the adjacency matrix can be regarded as a generalization of the weight enumerator of block codes (see the paragraph
right after Definition~\ref{D-Lambda}) this result comes somewhat surprising.
Indeed, there exist block codes that have the same weight enumerator but are not monomially equivalent; see Example~\ref{E-Forneyzero}(a)
at the end of this section.
This shows that the positivity of the Forney indices is certainly a necessary condition for the above result to be true.
On the other hand for block codes the famous MacWilliams' Equivalence Theorem~\cite{MacW63} tells us that isometric block codes are
monomially equivalent, see also, e.~g., \cite[Thm.~7.9.4]{HP03}.
It is not known yet whether an analogous statement is true for convolutional codes with a
suitable notion of isometry.
We believe that the result above will be helpful for investigating this issue.
\\[1ex]
\noindent\begin{proof}
The only-if part has been proven in Proposition~\ref{P-meLambda}.
Thus let us assume that $\bar{\Lambda}(\cC)=\bar{\Lambda}(\cC')$.
Since the adjacency matrices have the same size, the two codes have the same degree, say~$\delta$.
Let $G,\,G'$ be any basic and reduced encoder matrices of~$\cC$ and~$\cC'$ and
$(A,B,C,D)$ and $(A',B',C',D')$ be the corresponding controller forms, respectively.
Then the two systems have order~$\delta$ and, according to Proposition~\ref{P-CCF}, they
form canonical minimal realizations of the codes~$\cC$ and~$\cC'$.
Let~$\Lambda$ and~$\Lambda'$ be the associated weight adjacency matrices.
By assumption there exist $\phi\in\AutF$ and $T\in GL_{\delta}(\F)$ such that
\[
  \Lambda'_{X,Y}=\Lambda_{\phi(X)T,\phi(Y)T}\text{ for all }(X,Y)\in\F^{2\delta}.
\]
1) We first eliminate the automorphism~$\phi$ from this identity.
To this end, consider the realization $(A'',B'',C'',D'')=(\phi(A'),\phi(B'),\phi(C'),\phi(D'))$.
By Proposition~\ref{P-meLambda} this is a canonical realization of the code $\cC'':=\im\phi(G')$ and
this code is, by definition, monomially equivalent to~$\cC'$.
Moreover, by the form of~$A'$ and~$B'$ in the controller form we have $A''=A',\,B''=B'$.
Hence $(A',B',C'',D'')$ is a controller form.
Again by Proposition~\ref{P-meLambda} the associated weight adjacency matrix~$\Lambda''$ satisfies
\[
   \Lambda''_{\phi(X),\phi(Y)}=\Lambda'_{X,Y}=\Lambda_{\phi(X)T,\phi(Y)T}
   \text{ for all }(X,Y)\in\F^{2\delta}.
\]
Since~$\phi$ is a bijection on~$\F^{\delta}$ this yields
\begin{equation}\label{e-Lambda2}
   \Lambda''_{X,Y}=\Lambda_{XT,YT}\text{ for all }(X,Y)\in\F^{2\delta}.
\end{equation}
2) In \cite[Thm.~5.1]{GL05p} is has been proven that codes satisfying\eqnref{e-Lambda2}
have the same dimension and the same Forney indices.
Thus let $k=\dim(\cC)=\dim(\cC'')$.
Using Theorem~\ref{T-Lambdaunique} we may assume that both codes have their Forney indices in the same
ordering.
Let us denote them by $\nu_1\geq\ldots\geq\nu_k\geq1$.
Notice that $\delta=\sum_{i=1}^k\nu_i$.
Now the controller form implies $A'=A$ and $B'=B$.
Thus we arrive at canonical minimal realizations $(A,B,C,D)$ and $(A,B,C'',D'')$ such that the associated weight adjacency matrices satisfy\eqnref{e-Lambda2}.
It remains to show that the corresponding codes~$\cC$ and~$\cC''$ are monomially equivalent.
\\[.6ex]
3) We will show that
\begin{equation}\label{e-ABMU}
   A=T(A-MB)T^{-1}\text{ and }B=UBT^{-1}\text{ for some matrices }
   M\in \F^{\delta\times k},\,U\in GL_k(\F).
\end{equation}
By definition of the weight adjacency matrix we have for any $(X,Y)\in\F^{2\delta}$
\[
  Y-XA\in\im B\Longleftrightarrow \Lambda''_{X,Y}\not=0
  \Longleftrightarrow\Lambda_{XT,YT}\not=0\Longleftrightarrow
  YT-XTA\in\im B.
\]
Putting $\tilde{A}=TAT^{-1},\,\tilde{B}=BT^{-1}$, we thus get
\[
  Y-XA\in\im B\Longleftrightarrow Y-X\tilde{A}\in\im\tilde{B}.
\]
Using $X=0$ this implies $\im\tilde{B}=\im B$ and hence
$BT^{-1}=\tilde{U}B$ for some $\tilde{U}\in GL_k(\F)$.
On the other hand, for each $X\in\F^{\delta}$ there exists $u\in\F^k$ and $Y\in\F^{\delta}$ such that
$Y-XA=u B$, hence there exists $\tilde{u}\in\F^k$ such that $Y-X\tilde{A}=\tilde{u}B$.
This implies $X(\tilde{A}-A)=(u-\tilde{u})B$.
Using for~$X$ all standard basis vectors we obtain the identity $\tilde{A}=A+\tilde{M}B$ for some matrix $\tilde{M}\in\F^{\delta\times k}$.
Hence we arrive at $A=T^{-1}(A+\tilde{M}B)T$ and $B=\tilde{U}BT$.
This in turn yields\eqnref{e-ABMU}.
\\[.6ex]
4) In this step we will prove that $(A,B,C'',D'')$ and $(A,B,C,D)$ are related via the full feedback group followed by monomial equivalence.
Using Theorem~\ref{T-feedbackequiv} this will then establish the desired result.
In order to do so we will compare the entries of the weight adjacency matrices.
Consider the canonical minimal realization
$(\bar{A},\bar{B},\bar{C},\bar{D})=(TAT^{-1},BT^{-1},TC,D)$ of the code~$\cC$.
It is easy to see \cite[Rem.~3.6]{GL05p} that the associated weight adjacency matrix~$\bar{\Lambda}$ satisfies $\bar{\Lambda}_{X,Y}=\Lambda_{XT,YT}$ for all $(X,Y)\in\F^{2\delta}$ and hence Equation\eqnref{e-Lambda2}
implies
\[
   \bar{\Lambda}=\Lambda''.
\]
Now we can study the entries of these weight adjacency matrices.
Recall that~$\Lambda''$ belongs to the realization $(A,B,C'',D'')$ of the code~$\cC''$.
Since all Forney indices are positive, the matrix~$B$ has full rank~$k$ (see the controller form).
As a consequence, for each pair of states $(X,Y)\in\F^{2\delta}$ the set $\{XC''+uD''\mid u\in\F^k:\, Y=XA+uB\}$ has at most one element.
Recalling the definition of the weight adjacency matrix in\eqnref{e-LambdaXY} one obtains that the nonzero entries are given by
\begin{equation}\label{e-LambdaId}
  \Lambda''_{X,XA+uB}=\bar{\Lambda}_{X,XA+uB}\text{ for all }(X,u)\in\F^{\delta}\times\F^k,
\end{equation}
and these entries have the value $\Lambda''_{X,XA+uB}=W^{\alpha}$ where
$\alpha=\wt(XC''+uD'')$.
On the other hand notice that, due to\eqnref{e-ABMU}, for any $(X,u)\in\F^{\delta}\times\F^k$ we have
\[
  XA+uB=X(TAT^{-1}-TMBT^{-1})+uUBT^{-1}=X\bar{A}+\bar{u}\bar{B}\text{ where }\bar{u}=uU-XTM.
\]
Thus\eqnref{e-LambdaXY} yields
$\bar{\Lambda}_{X,XA+uB}=\bar{\Lambda}_{X,X\bar{A}+\bar{u}\bar{B}}=W^{\beta}$ where
$\beta=\wt(X\bar{C}+\bar{u}\bar{D})$.
As a consequence,\eqnref{e-LambdaId} implies
\[
  \wt\bigg((X,u)\begin{pmatrix}C''\\D''\end{pmatrix}\bigg)=
  \wt\big(X\bar{C}+(uU-XTM)\bar{D}\big)
  =\wt\bigg((X,u)\begin{pmatrix}\bar{C}-TM\bar{D}\\U\bar{D}\end{pmatrix}\bigg)
\]
for all $(X,u)\in\F^{\delta}\times\F^k$.
Now \cite[Lemma~5.4]{GL05p}, which is basically MacWilliams' Equivalence Theorem for block codes, yields the existence of a permutation matrix~$P\in GL_n(\F)$ and a nonsingular diagonal matrix~$R\in GL_n(\F)$ such that
\[
  \begin{pmatrix}C''\\D''\end{pmatrix}=
  \begin{pmatrix}\bar{C}-TM\bar{D}\\U\bar{D}\end{pmatrix}PR.
\]
Hence the realization $(A,B,C'',D'')$ of~$\cC''$ is of the form
\begin{align*}
   (A,B,C'',D'')&=(T(A-MB)T^{-1},UBT^{-1},(\bar{C}-TM\bar{D})PR,U\bar{D}PR)\\
    &=(T(A-MB)T^{-1},UBT^{-1},T(C-MD)PR,UDPR).
\end{align*}
This finally allows us to apply Theorem~\ref{T-feedbackequiv}, which then tells us that
\[
    \cC''=\im\big(B(z^{-1}I-A)^{-1}C''+D''\big)=\im\big(B(z^{-1}I-A)^{-1}C+D\big)PR=\im(GPR)
\]
is monomially equivalent to~$\cC$.
This completes the proof.
\end{proof}

\begin{rem}\label{R-strongmonomequiv}
The proof shows that the result of Theorem~\ref{T-me} is also true if monomial equivalence of codes and equivalence of adjacency matrices (see Definitions~\ref{D-me} and~\ref{D-Lambdaequiv}) do not allow nontrivial field automorphisms~$\phi$.
In that case step~1) is simply omitted.
\end{rem}

We close the paper with presenting some examples showing that the theorem above is not true if some of the Forney indices are zero.

\begin{exa}\label{E-Forneyzero}\
\begin{alphalist}
\item Recall that for a block code $\cC=\im G$, thus $G\in\F^{k\times n}$, the
      adjacency matrix is the ordinary weight enumerator.
      In this case it is well known that block codes with the same weight enumerator are,
      in general, not monomially equivalent. The following example is taken from \cite[Exa.~1.6.1]{HP03}.
      The matrices
      \[
         G_1=\begin{pmatrix}1&1&0&0&0&0\\0&0&1&1&0&0\\1&1&1&1&1&1\end{pmatrix},\
         G_2=\begin{pmatrix}1&1&0&0&0&0\\1&0&1&0&0&0\\1&1&1&1&1&1\end{pmatrix}
         \in\F_2^{3\times 6}
       \]
       generate codes with the same weight enumerator $1+3W^2+3W^4+W^6$, but are not monomially equivalent.
       The latter follows from $G_1G_1\T=0\not= G_2G_2\T$.
\item From the previous data one can also construct an example with
      positive degree.
      Using the rows of the matrices above in a suitable way one obtains
      \[
         G=\begin{pmatrix}1&1&z&z&0&0\\1&1&1&1&1&1\end{pmatrix},\quad
         \bar{G}=\begin{pmatrix}z+1&1&z&0&0&0\\1&1&1&1&1&1\end{pmatrix}\in\F_2[z]^{2\times 6}.
      \]
      Both matrices are basic and reduced.
      The weight adjacency matrices of the associated controller forms are both given by
      \[
         \Lambda=\begin{pmatrix}1+W^6&W^2+W^4\\W^2+W^4&W^2+W^4\end{pmatrix}.
      \]

      But the codes $\cC=\im G$ and $\bar{\cC}=\im\bar{G}$ are not monomially equivalent.
      This can be seen by computing $UG$ for all $U\in GL_2(\F_2[z])$ such that $UG$ is reduced with indices~$1$ and~$0$ again. The only options are
      \[
         U\in\Big\{I_2,\,\begin{pmatrix}1&1\\0&1\end{pmatrix},\,
       \begin{pmatrix}1&z\\0&1\end{pmatrix},\,\begin{pmatrix}1&1+z\\0&1\end{pmatrix}\Big\}
      \]
      and it is seen by inspection that in none of these cases $UG$ has, up to ordering, the same columns as~$\bar{G}$.
\end{alphalist}
\end{exa}

\section*{Conclusion}
In this note we have shown that codes with all Forney indices being positive are monomially
equivalent if and only if they share the same adjacency matrix.
The result is not true for codes with at least one Forney index being zero (unless they are one-dimensional block codes).
We believe that this result will be helpful for the investigation of ($\F$-linear) isometries for convolutional codes that leave the characteristic properties of the codes invariant.
This subject however has to remain open for future research since no
well founded notion for such isometries has been established yet.
Once this has been found the question of a MacWilliams' Equivalence Theorem for
convolutional codes can be addressed, and our result might provide a partial answer.

\bibliographystyle{abbrv}
\bibliography{literatureAK,literatureLZ}

\end{document}